\documentclass[conference]{IEEEtran}

\usepackage{graphicx}
\usepackage{amsmath}

\ifCLASSINFOpdf
\else
\fi
\begin{document}
%
\title{Modeling the Annual Growth Rate of Electricity Consumption of China in the 21st Century: \newline Trends and Prediction}

\author{\IEEEauthorblockN{CHEN Kunjin}
\IEEEauthorblockA{Department of Electrical Engineering\\
Tsinghua University\\
Beijing, 100084, P. R. China\\
Email: yalickj@163.com}
\and
\IEEEauthorblockN{CHEN Kunlong}
\IEEEauthorblockA{Department of Electrical Engineering\\
Beijing Jiaotong University\\
Beijing, 100044, P. R. China\\
Email: 15121396@bjtu.edu.cn}}


%


\maketitle
\begin{abstract}
In this paper, the annual growth rate of electricity consumption (EC) of China in the first 15 years of the 21st century is modeled using multiple linear regression (MLR). Historical data and trends of gross domestic product (GDP), fixed assets investment (FAI) and share of secondary sector in China's GDP are used to account for the observed trend and fluctuation of EC in recent years. A comparison between the proposed method and the predictions given by China Electricity Council is performed, showing that the proposed method is good at capturing the trend of the EC of China under complex social and economic background. The ECs for 2016 and 2017 are also predicted, which is helpful for energy planners and policy makers for future challenges.
\end{abstract}

\smallskip


%
\IEEEpeerreviewmaketitle

\section{Introduction}

Electricity demand forecasting has been a hot topic in the past few years not only because it provides information for power system investment planning, but also due to the growing concern over environment issues on a global basis. The first fifeteen years of the 21st century has witnessed the stunning economic growth of China, and the ever-growing electricity consumption (EC) of this big developing country has helped make it the largest carbon emitter in the world, imposing heavy strains on aspects of energy conservation and emission reduction. Therefore, it is of great interest to find a way to predict the trend of EC of China with high accuracy. Several factors magnifies the challenge of correctly predicting the EC of China in the near future. Firtly, the domestic economic downturn and the unfavorable global economic environment would inevitably decelerate the growth of EC. In addition, as the economic structure of China continue to transform, the investment-driven growth is deemed to be unsustainable, and the EC may start to decouple from certain quantities including gross domestic product (GDP). Further, the energy structure of China causes difficulty for pollution control and Carbon emission reduction, by which the generation of electricity is directly affected.

A lot of researches of EC prediction have been carried out using a variety of methods, such as regression analysis \cite{bianco2009electricity}, time series analysis \cite{li2013hybrid}, grey prediction methods \cite{akay2007grey}, Artificial neural network \cite{Azadeh2008Annual}, etc. In this paper, we aim to propose a novel approach to model the annual growth rate (AGR) of EC of China in the 21st century by using Historical data of GDP, fixed assets investment (FAI) and the share of the secondary sector in China’s GDP, as these quantities can reflect the development phase and practical conditions of China. Our research proves that these predictors can effectively fit the data, on which basis we move on to predict the EC of China in 2016 and 2017.

\section{Method}
\subsection{Multiple Linear Regression {\rm \cite{james2013introduction}}}
MLR is an approach for modeling the relationship between a variable $Y$ and several variables (or predictors) denoted by $X$. Oftentimes, the MLR model takes the form 
\begin{equation}
Y=\beta_0+\beta_1X_1+\beta_2X_2+\cdots+\beta_pX_p+\epsilon
\end{equation}
where $X_j$ represents the jth predictor and $\beta_j$ quantifies the association between that variable and the response. The least squares approach is often used to estimate the parameters. We estimate $\hat{\beta_1},\hat{\beta_2},...\hat{\beta_n}$ to minimize the sum of squared residuals. More specifically, the objective of the approach can be written as 
\begin{equation}
{\rm minimize }\sum_{i=1}^n(y_i-\hat{\beta_0}-\hat{\beta_1}x_{i1}-\hat{\beta_2}x_{i2}-\cdots-\hat{\beta_p}x_{ip})^2
\end{equation}
Not every relationship between $Y$ and $X$ that we meet can be described by a linear model. In order to assess the validity of this method, it’s necessary to test the heteroskedasticity and normality of the fitting deviation, for the reason that we assume $\epsilon$ obeys a gaussian distribution and the predictors are linear independent. In this paper, Breusch-Pagan test is used for testing the heteroskedasticity and Shapiro-Wilk test is used for testing the normality.

The main advantage of the linear regression method is that it is not prone to the problem of over-fitting when data is insufficient due to its lower model flexibility compared with complex nonlinear models. Besides, the result of linear regression can be used to explain the relationship between $Y$ and every single explanatory variables $X_j$. This can be quite useful when future analysis with economic application is conducted.

\subsection{Critical Factors Accounting for Recent Trends of Annual Growth Rates of Electricity Consumption}
As GDP is a good measure of economic activity, it is reasonable to infer that a positive correlativity can be found between the EC and GDP of a country. The AGRs of GDP and EC from the year 1998 to the year 2014 are plotted in Fig. 1. In spite of the fact that the trend of AGRs of EC is highly correlated to the AGRs of GDP, it is observed that the former has greater fluctuation. Thus, in order to implement the MLR model, we need to find other factors affecting the EC of China.

A variety of factors may account for such a phenomenon. The first quantity of interest is FAI, which has been proven to have long-term impact on the economy \cite{podrecca2001fixed}. Considering the development stage and economic structure of China, it is likely that the trend of FAI influences the growth of EC in the long run. Note that the majority of the factors considered in the literature have instant effects on EC (in other words, to model the quantity of EC of a year, the quantities of the factors of this specific year is required). The effect of FAI on the AGRs of EC, however, is expected to be delayed rather than instant. In Fig. 1, we illustrate the delayed effect of the AGRs of FAI on the AGRs of EC with markers ``1'', ``2'', ``3'', ``4'' and colored areas corresponding to the markers. Each marker indicates a turn in the trend of the AGRs of FAI. When the trend turns upwards (i.e., 1999-2000 and 2006-2007), the AGRs of EC are expected to exceed the AGRs of GDP during three to five years after the turning point, and the reverse is also true (the year 2003-2004 and the year 2009-2010 are the turning points).

\begin{figure}[b]
\centering
\includegraphics[height = 6.5cm, ]{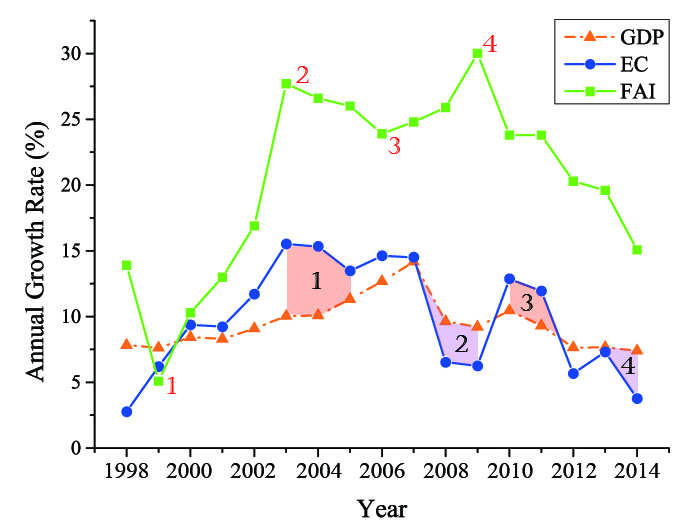}
\caption{AGRs of GDP, EC and FAI from 1998 to 2014. Four colored areas between the lines of GDP and EC correspond to the markers indicating trend turning points of the AGRs of FAI.}
\label{fig:1} 
\end{figure}

To highlight the ``turning point'' effect of AGRs of FAI, the FAI indicator $I$ is introduced in this paper. Given the AGRs of FAI of the current year and previous two years ($Y_{n}$, $Y_{n-1}$ and $Y_{n-2}$), the FAI indicator $I_{n}$ is calculated as
\begin{equation}
\begin{split}
I_n=Y_n-2(Y_{n-1}-Y_{n-2})\\=Y_n+2Y_{n-2}-2Y_{n-1}
\end{split}
\end{equation}
Concretely, in order to investigate the impact of the ``turning point'' effect up to five years, $I_{n-1}$, $I_{n-2}$, $I_{n-3}$ and $I_{n-4}$ are considered for the year indicated by $n$.

The economic structure also has a significant impact on the EC, as the electricity demands of different sectors in the economy are not the same. For instance, energy intensive industries use more electricity than knowledge intensive service industries per unit of output. In order to take into account the impact of the change of economic structure (especially the fact that the proportion of the secondary sector is continuously decreasing), the share of the secondary sector in China’s GDP is used. In this paper, we denote the share of the secondary sector of the current year and the previous year as $S_{n}$ and $S_{n-1}$, and the change of these two quantities are calculated and denoted as $D_{n}$, that is,
\begin{equation}
D_{n}=S_{n}-S_{n-1}
\end{equation}

\section{Results and discussion}
\subsection{Fitting Results of the MLR Model}

Historical data from 2002 to 2013 is used for the MLR, and regression through the origin (used when economic theory suggests hat $\beta_0$ is zero \cite{wooldridge2015introductory}) is implemented in this paper. Concretely, we consider the baseline situation with all economic and social indexes of a certain year being the same for five consecutive years. This obviously indicates that the independent varibles used in this paper are all zero. Thus, we make the assumption that when all the independent variables are zero, the EC of the year is indentical to that of the baseline situation, which requires a zero $\beta_0$. After eliminating the insignificant variables, the final MLR model we obtain is
\begin{equation}
\hat{C}=1.11904\times G_{n} + 0.17232\times I_{n-4} + 2.53056\times D_{n} 
\end{equation}
where $\hat{C}$ is the predicted EC for the year denoted by $n$. The adjusted $R^2$ of the model is 0.975 and the p-values are $5.84\times10^{-9}$, $0.0473$ and $0.0039$ for $G_{n}$, $I_{n-4}$ and $D_{n}$, respectively. The actual and fitted AGRs of EC is plotted in Fig. 2.

\begin{figure}[t]
\centering
\includegraphics[height = 6.5cm, ]{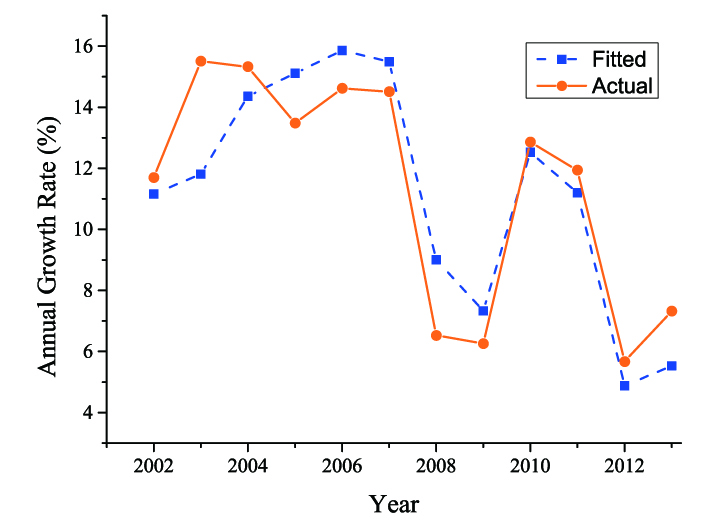}
\caption{ Fitted and actual AGRs of EC from 2002 to 2013. The fitted AGRs are obtained by the MLR model using data from 2002 to 2013.}
\label{fig:2} 
\end{figure}

The fitted and actual AGRs and the errors of the MLR model using data from 2002 to 2013 are listed in Table I. The errors are calculated using the quantities of annual EC instead of comparing the fitted and actual AGRs. The mean absolute percentage error (MAPE) of the years from 2002 to 2013 is 1.210, which indicates that the MLR model is quite effective.

\begin{table}[t]
\centering  
\caption{Fitting Results and Errors of the \protect\\ MLR Model (2002 to 2013) }
\begin{tabular}{cccc}  
\hline
Year & Fitted (\%) & Actual (\%) & Error  (\%)\\ 
\hline  
2002&11.163&	11.696&	$-$0.477\\
2003&11.815&	15.508&	$-$3.197\\
2004&14.362&	15.326&	$-$0.836\\
2005&15.110&	13.475&	1.441\\
2006&15.864&	14.619&	1.086\\
2007&15.491&	14.510&	0.857\\
2008&9.004&	6.528&	2.324\\
2009&7.331&	6.262&	1.005\\
2010&12.524&	12.860&	$-$0.298\\
2011&11.202&	11.939&	$-$0.657\\
2012&4.875&       5.675&	$-$0.756\\
2013&5.532&	7.324&	$-$1.669\\
\hline
\end{tabular}
\end{table}

The heteroskedasticity test and normality test are carried out after fitting the model, and the results are listed in Table II. The p-values of these two methods are larger than the critical value, which proves the validity of the model.

\begin{table}[!t]
\centering  
\caption{Results of Regression Diagnostics}
\begin{tabular}{ccc}  
\hline
Method & Critical value  & P-value \\ 
\hline  
Breusch-Pagan test  &   0.05       &           0.80\\
Shapiro-Wilk test   &   0.05             &     0.56 \\

\hline
\end{tabular}
\end{table}

\subsection{Performance of the MLR Model}
We use the data of 2014 and 2015 to validate the proposed method. As the AGR of GDP of China is usually close to the predicted value given by the National Bureau of Statistics of China (NBSC), the AGRs of GDP predicted by the NBSC are used directly. In addition, the FAI indicators of four years in advance are readily available. Thus, to ``predict'' the AGRs of EC in 2014 and 2015, we only need to obtain the GDP shares of the secondary sector in these two years using historical data. The contribution of the secondary sector to the GDP of China (share of GDP) from 2006 to 2013 is shown in Fig. 3.

\begin{figure}[t]
\centering
\includegraphics[height = 6.5cm, ]{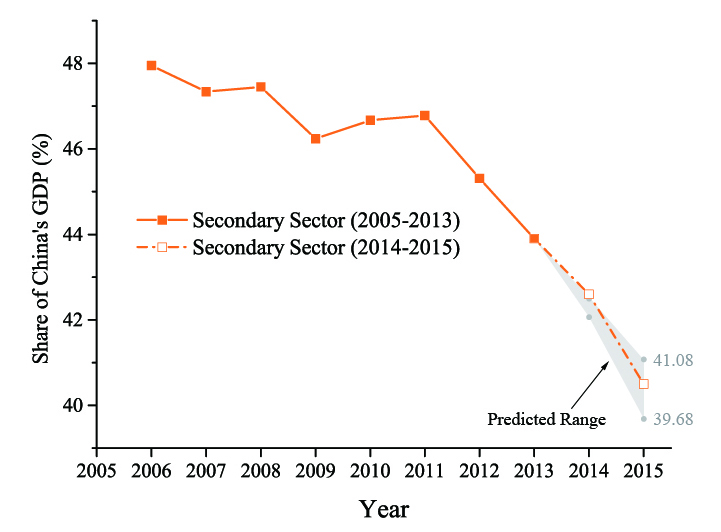}
\caption{The share of secondary sector in China’s GDP from 2005 to 2015. The gray area indicates the predicted range for 2014 and 2015.}
\label{fig:3} 
\end{figure}

The traditional three-sector theory developed by Fisher, Clark and Fourasti\'e points out the fact that the economic activity of a country shifts from the primary sector to the tertiary sector through the secondary sector. Based on the theory, the share of the secondary sector is expected to decrease upon the economy enters the de-industrialization stage, and the trend may accelerate for several years as the tertiary sector experiences booming development. Recent observations on early advancement of the tertiary sector of developing countries also support this expectation \cite{lutz2013characterizing}. Specifically, the share of the tertiary sector of China exceeded the share of the secondary sector in 2013 with an accelerated trend. Though the observed trend of China is believed to be the consequence of active adjustment of industrial structure rather than de-industrialization, the impact of the trend on the AGRs of EC is still effective. Thus, in this paper, we ``predict'' the share of the secondary sector based on the assumption that the contribution of the secondary sector to the GDP of China declines continuously. A predicted range is estimated by giving an upper bound and a lower bound of prediction. Given the share of GDP of previous two years, $S_{n-1}$ and $S_{n-2}$, the upper bound of the predicted share of GDP of the current year, $S_n^h$, is calculated as
\begin{equation}
\begin{split}
S_n^h=S_{n-1}+(S_{n-1}-S_{n-2})\\=2S_{n-1}-S_{n-2}
\end{split}
\end{equation}
The lower bound of the predicted share of GDP of the current year is similarly calculated as
\begin{equation}
\begin{split}
S_n^l=S_{n-1}+1.3(S_{n-1}-S_{n-2})\\=2.3S_{n-1}-1.3S_{n-2}
\end{split}
\end{equation}
where the accelerating coefficient of the term $S_{n-1}-S_{n-2}$ changes to 1.3. This corresponds to the accelerated downtrend. The predicted ranges of the year 2014 and 2015 is calculated using (4) and (5). In Fig. 3, the gray area indicates the predicted ranges for 2014 and 2015. Actual shares of the secondary sector for 2014 and 2015 are also given in Fig. 3. For the year 2014, the actual share of the secondary sector slightly exceeds the upper bound of the predicted range. The actual share of the secondary sector of the year 2015 is found near the middle of the predicted interval.

The predicted AGRs and errors of 2014 are listed in Table III. For the proposed method, the AGR of GDP is the predicted value given by NBSC at the beginning of 2014, which is 7.5\%. In addition, the $I_{n-4}$ for 2014 is $-$0.103 and the predicted range of $D_{n}$ for 2014 is $-$1.41\% to $-$1.83\%. The China Electricity Council (CEC) also predicts the AGR of EC at the beginning of each year, and the predicted value for 2014 is 7.0\%. It is clear from the table that both the low and high values given by the proposed method are closer to the actual AGR of EC of 2014 than the value given by CEC.

\begin{table}[b]
\centering  
\caption{Annual Growth Rates and Errors of the Year 2014}
\begin{tabular}{ccc}  
\hline
Methods & Annual Growth Rates (\%) & Error (\%)\\ 
\hline  

The proposed method (low) &   1.979 & $-$1.731\\
The proposed method (high) &  3.050 & $-$0.700\\
CEC & 7.000 & 3.106\\
Actual & 3.777 & / \\

\hline
\end{tabular}
\end{table}

Table IV lists the predicted AGRs and errors of 2015. For the proposed method, the AGR of GDP given by NBSC at the beginning of 2015 is 7.0\%, the $I_{n-4}$ for 2014 is 0.062 and the predicted range of $D_{n}$ for 2014 is $-$1.52\% to $-$2.92\%. Moreover, the predicted AGR given by CEC is 4.5\%. As we can see, the actual AGR of EC in 2015 is lower than all predictions with an AGR very close to 0. The actual AGR also lies out of the predicted range of the proposed method, though the lower bound of the proposed method achieves an error of 0.923$\%$, which is satisfactory. Several reasons may have contributed to the relatively low AGR. The first is the ``supply side reforms'' carried out in the energy-intensive industries as a result of the severe industrial overcapacity. Further, as the economic structure keeps on evolving, the growth of EC tends to be more stable than in the past.   

\begin{table}[b]
\centering  
\caption{Annual Growth Rates and Errors of the Year 2015}
\begin{tabular}{ccc}  
\hline
Methods & Annual Growth Rates (\%) & Error (\%)\\ 
\hline  

The proposed method (low) &   1.523 & 1.034\\
The proposed method (high) &  5.056 & 4.549\\
CEC & 4.500 & 3.997\\
Actual & 0.483 & / \\

\hline
\end{tabular}
\end{table}

\subsection{Predicting the EC of 2016 and 2017}
We use data from 2004 to 2015 to predict the EC of 2016 and 2017. The MLR model we obtain for this period of time is
\begin{equation}
\hat{C}=1.06443\times G_{n} + 0.22177\times I_{n-4} + 2.51926\times D_{n} 
\end{equation}
The adjusted $R^2$ of the model is 0.977 and the p-values are $4.16\times10^{-9}$, $0.0101$ and $0.0003$ for $G_{n}$, $I_{n-4}$ and $D_{n}$, respectively.
 
The expected AGR of GDP for 2016 given by NBSC is 6.5\% to 7.0\%, and the $I_{n-4}$ is $-$0.035. Moreover, based on (6) and (7), the predicted range of $D_{n}$ for 2016 is $-$2.73\% to $-$2.10\%. Therefore, the lower and higher bounds of the predicted AGR of EC are $-$0.7\% and 0.8\%, respectively. As the EC of China in 2015 is 5.550 trillion kWh, the predicted EC for 2016 is between 5.511 trillion kWh and 5.594 trillion kWh. Further, if we fix the expected AGR of GDP for 2017 to 6.5\%, then the predicted lower and higher bounds of the AGR of EC in 2017 are $-$1.4\% and 2.2\%.

\section{Conclusion and Future Outlook}
In this paper, MLR is used to model the AGRs of EC of China. Historical data of GDP, FAI and economic structure is used to construct the model. Results show that the performance of the MLR model is favorable. Concretely, the instant effects of AGRs of GDP and the share of the secondary sector in China's GDP as well as the delayed effect of FAI are considered to account for the observed trend and fluctuation of EC. The predicted ECs of 2016 and 2017 indicate that the demand of electricity will remain stable.

As the development pattern and economy structure of China keeps on transforming, the remarkable high-speed growth of EC may come to an end. Under the background of global climate change and global warming, the active transform of energy structure is indispensable to the sustainable development of China as well as the world. Therefore, it is now of higher priority to focus on improving the quality of electricity supply thoughout the ``life cycle'' of the generated elelctricity. The dramatic changes in the pattern of both source end (e.g., renewable energy integration) and consumption end (e.g., the penetration of electric vehicles) are expected to have profound impacts on the evolvement of the power grid and other energy supply chains of China. Hence, solutions such as ``energy internet'' and ``smart grid'' are promising to help shape the future landscape of energy, as they are capable of improving efficiency and customer side experiecne as well as promoting cyber-energy integration and market participation. It is believed that the method presented in this paper would help planners in the energy industry and policy makers to address future challenges.





%



\bibliographystyle{IEEEtran}%
\bibliography{UCGES2016.bib}

\end{document}